\documentclass[journal,twoside,web]{ieeecolor2}
\usepackage{generic}
\usepackage{cite}
\usepackage{amsmath,amssymb,amsfonts}
\usepackage{algorithmic}
\usepackage{graphicx}
\usepackage[detect-all=true, separate-uncertainty=true, multi-part-units=single,binary-units=true]{siunitx}
\usepackage{textcomp}
\usepackage{background}
\backgroundsetup{
  position=current page.south,
  angle=0,
  nodeanchor=south,
  vshift=10pt,
  opacity=0.6,
  scale=2,
  contents={\textbf{Accepted for publication in IEEE Sensors Journal}}
}

\usepackage{pdfpages} % <— key package
\usepackage{hyperref}

\usepackage{subcaption}
\usepackage[T1]{fontenc}
\usepackage[utf8]{inputenc}
\usepackage{newunicodechar}
\newunicodechar{：}{:}
\DeclareUnicodeCharacter{03BC}{\ensuremath{\mu}}
\usepackage{svg}
\usepackage{xcolor}

\def\BibTeX{{\rm B\kern-.05em{\sc i\kern-.025em b}\kern-.08em
    T\kern-.1667em\lower.7ex\hbox{E}\kern-.125emX}}
\begin{document}
\title{Volumetric ultrasound imaging with a sparse matrix array and integrated fiber-optic sensing for robust needle tracking in interventional procedures}

\author{
    Weidong Liang$^{1,*}$,  
    Javad Rostami$^{1,*}$,  
    Christian Baker$^{1}$,  
    Simeon West$^{2}$, Athanasios Diamantopoulos$^{3}$,  
    Sunish Mathews$^{4}$,  
    Adrien E. Desjardins$^{5}$,  
    Sebastien Ourselin$^{1}$,  
    Laura Peralta$^{1}$,  
    and Wenfeng Xia$^{1,\ddagger}$  
    \thanks{*These authors contributed equally to this work.}  
\thanks{$^{\dagger}$Corresponding author: Wenfeng Xia (wenfeng.xia@kcl.ac.uk)}  
    \thanks{This research was funded in whole or in part by the Wellcome Trust (203148/Z/16/Z), EPSRC (NS/A000049/1), The Royal Society [URF/R1/211049], and the King's-China Scholarship Council PhD Scholarship program (K-CSC). For the purpose of Open Access, the author has applied a CC BY public copyright license to any Author Accepted Manuscript version arising from this submission.}  
    \thanks{$^{1}$ School of Biomedical Engineering \& Imaging Sciences, King's College London, London, United Kingdom.}  
    \thanks{$^{2}$ Department of Anaesthesia, University College London Hospital, London, United Kingdom.}  
    \thanks{$^{3}$ Department of Interventional Radiology, Guy's \& St Thomas' NHS Foundation Trust, London, United Kingdom.}
    \thanks{$^{4}$ School of Science and Engineering, University of Dundee, United Kingdom.}  
    \thanks{$^{5}$ Department of Electrical and Computer Engineering, The University of British Columbia (UBC), Vancouver, Canada.}  
}
\maketitle

\begin{abstract} 
Objective: Accurate visualization of interventional devices, such as medical needles, in relation to the procedural target is critical for the safe and effective guidance of interventional procedures. Ultrasound (US) imaging is widely used for guiding percutaneous needle interventions, but the 2D nature of most clinical US probes limits accurate 3D localization, particularly of the needle tip. In this work,  we introduce a novel system that combines both volumetric US imaging and 3D needle tracking. Method: The proposed novel system integrates a fiber-optic hydrophone (FOH) into the needle and employs a 2D sparse spiral US array with 256 active elements. Real-time volumetric US imaging was achieved by plane-wave imaging with the sparse array, while
3D needle tip tracking was enabled through communications between the US probe and the FOH. Results: The system achieved spatial resolutions (mean ± standard deviation) of 2.06 ± 0.29\,mm (lateral), 2.26 ± 0.23\,mm (elevational), and 0.69 ± 0.12\,mm (axial) at depths ranging from 10 to 40\,mm. The tracking accuracy was better than 0.30 ± 0.21\,mm. The clinical potential of the system was demonstrated using a nerve block training phantom. Conclusions: This study presents a proof-of-concept for an integrated solution that enables simultaneous volumetric anatomical imaging and precise 3D needle tip tracking. Significance: The proposed system holds strong potential to enhance the efficacy and safety of image-guided interventional procedures by providing real-time 3D anatomical visualization and accurate needle tracking.

\end{abstract}


\section{Introduction}
\label{sec_1}
In minimally invasive procedures such as fetal surgery, tumor biopsy, and regional anesthesia, it is crucial for clinicians to accurately visualize and locate intraoperative tools such as needles relative to patient anatomy and procedure targets ~\cite{chin2008needle, roy2017concurrent, kumar2004recent}. Ultrasound (US) imaging is often preferred over X-ray and magnetic resonance imaging (MRI), due to its advantages of non-ionizing radiation, portability, real-time imaging capability, and cost-effectiveness~\cite{o2010image}. However, most clinical US imaging systems provide only 2D views in real-time, posing a significant limitation for guiding interventions. In particular, accurately tracking and guiding the needle, particularly its tip, in 3D space towards the procedural target, while simultaneously avoiding critical tissue structures visible in 2D US images, imposes a significant cognitive load on clinicians. Moreover, during in-plane insertions, it can be challenging to clearly distinguish the needle from the background with needle insertions into highly echogenic tissues (e.g., adipose tissue). \cite{reusz2014needle}. Needle visibility is also worsened with steep insertion angles due to strong specular reflections at the smooth surface of the needle~\cite{chin2008needle}. With out-of-plane insertions, the intersection of the needle shaft with the imaging plane may be mistaken for the needle tip~\cite{course2023ultrasound}. These challenges not only elevate the risk of procedural complications, such as bleeding, infection, tissue damage, and pregnancy loss~\cite{bhatia2010pneumothorax,o2010image}, but also place substantial demands on the clinician's technical skill and cognitive focus~\cite{chin2008needle}. 

%Poor needle visualization, particularly at the needle tip, increases the risk of procedural failure and can lead to complications such as bleeding, infection, tissue damage, and pregnancy loss \cite{bhatia2010pneumothorax,o2010image}. 

Various solutions have been proposed to enhance US-guided needle procedures, with several now in use in clinical practice~\cite{sato2024basic, hovgesen2022echogenic,beigi2021enhancement, hebard2011echogenic}. For example, in regional anesthesia, clinicians might use a needle puncture guide mounted onto the US probe, which provides mechanically fixed insertion paths to simplify user hand-eye coordination. However, this mechanism may restrict free-hand maneuverability and prove less effective for skilled clinicians\cite{sato2024basic}. Another prominent approach uses needles with echogenic coatings designed to enhance visibility under US imaging~\cite{hovgesen2022echogenic, beigi2021enhancement}. Echogenic needles can improve visibility, particularly at steep insertion angles\cite{hebard2011echogenic}. However, they tend to be costly, and it is still not feasible to track the needle tip when it operates out of the imaging plane. 

Efforts have also been directed towards 3D US imaging to provide clinicians with volumetric visibility of patient anatomy and interventional devices~\cite{huang2017review}. One method involves free-hand scanning combined with position sensors, which enables 3D reconstruction by spatially registering 2D-image slices using real-time tracking data~\cite{nelson1998three}. Alternatively, mechanically motorized probes can acquire 3D data through controlled movements such as rotation or tilting~\cite{prager2010three}. However, these methods are limited by prolonged image acquisition times, mechanical complexity, and reduced flexibility~\cite{huang2017review}. While matrix imaging arrays support real-time volumetric US with high frame rates, they demand high-channel-count systems and considerable computational resources, making them costly to implement. These challenges can be mitigated by using sparse array probes~\cite{lookwood1996optimizing, ramalli2015density}. Nevertheless, implementing sparse array methodologies in 3D medical imaging and image-guided minimally invasive interventions remains largely underexplored. 

In parallel, other studies have explored alternative imaging methods to enhance needle visibility, including Power Doppler US~\cite{daoud2018accurate,daoud2022needle}, and photoacoustic imaging~\cite{shi2022improving, shi2024photoacoustic}. Power Doppler methods visualize needles vibrated by an integrated piezoelectric actuator. However, their performance can be significantly affected by variations in the speed of sound and tissue density. Photoacoustic imaging provides another promising solution, as the strong optical absorption of metallic needles produces high-contrast signals. Nevertheless, needle tracking using photoacoustic techniques has generally been restricted to 2D due to system complexity and integration challenges~\cite{shi2024photoacoustic}. Electromagnetic (EM) sensor-integrated needles offer an alternative approach by detecting external EM fields generated by a nearby field generator, enabling 3D needle tracking~\cite{reusz2014needle, fenster2000three, seitel2024miniaturized}. However, EM tracking systems are prone to interference from surrounding metallic instruments, leading to field distortion and reduced accuracy. Moreover, the requirement for a relatively bulky field generator positioned near the clinical site poses practical challenges and can deter clinical adoption. 

Ultrasonic needle tracking has emerged as a compelling alternative to EM tracking. This technique employs a miniaturized US transducer integrated at the needle tip to transmit and/or receive US signals, facilitating real-time interaction with an external US probe for accurate needle tracking~\cite{breyer1984ultrasonically,baker2022intraoperative}. Early works included the use of an integrated active piezoelectric transducer\cite{breyer1984ultrasonically}. In 2011, Mung et al.~\cite{mung2011non} presented an active tracking system that used a piezoelectric transmitter embedded in the catheter tip, which was reported to achieve sub-millimeter accuracy. In 2015, Guo et al.~\cite{guo2015integrated} introduced an active needle-integrated transponder capable of both receiving and transmitting US signals at the needle tip, enabling tip localization as a bright echo spot within the US image. However, piezoelectric transducers suffer from inherent performance limitations that hinder their effectiveness in miniaturized applications. These include high manufacturing costs, significant sensitivity and transmission power losses as the transducer surface area decreases, limited bandwidth, and strong directional dependence. Moreover, achieving adequate signal generation often requires high-voltage excitation, potentially posing a risk to patients, particularly during sensitive procedures such as cardiac interventions.

To address these challenges, fiber-optic US transducers have emerged as a promising approach for needle tracking, offering acoustic sensing or transmission via thin optical fibers. Compared to conventional piezoelectric transducers, fiber-optic technologies provide several advantages, including superior miniaturization, omnidirectional sensitivity, broader bandwidth, and improved cost-effectiveness. Recent studies have explored using fiber-optic transmitters (FOT) embedded within needles, which generate US signals via the photoacoustic effect for active tip tracking~\cite{xia2017ultrasonic}. However, these studies were limited to 2D tracking. In our previous works \cite{xia2015plane, xia2016coded, baker2022intraoperative}, we proposed a needle tip tracking system based on a fiber-optic hydrophone (FOH) integrated with a clinical 2D US imaging system. The FOH, featuring a Fabry-Pérot (FP) cavity, was miniaturized and integrated into a 20-gauge needle, with the sensing element aligned to the needle tip. In 2017, Xia et al. \cite{xia2017looking} also reported a 3D needle tracking system using a needle-integrated FOH with a specialized imaging probe featuring a central array for 2D US imaging and two additional arrays on each side of the imaging array dedicated to tracking. This system enabled out-of-plane needle tip tracking by detecting transmissions from the side arrays using the FOH. 

To date, most approaches have tackled the challenge of 3D intraoperative guidance during percutaneous needle procedures by focusing solely on either needle tip tracking or volumetric US imaging. In this study, we propose an integrated solution that seamlessly merges both capabilities into a single platform. Our system incorporates an FOH embedded at the needle tip for 3D tracking and utilizes a 2D sparse array to enable real-time 3D imaging, offering integrated needle tracking and anatomical visualization in a compact and efficient configuration.

\begin{figure*}[!ht]
    \centering
    \includegraphics[width=0.95\textwidth]{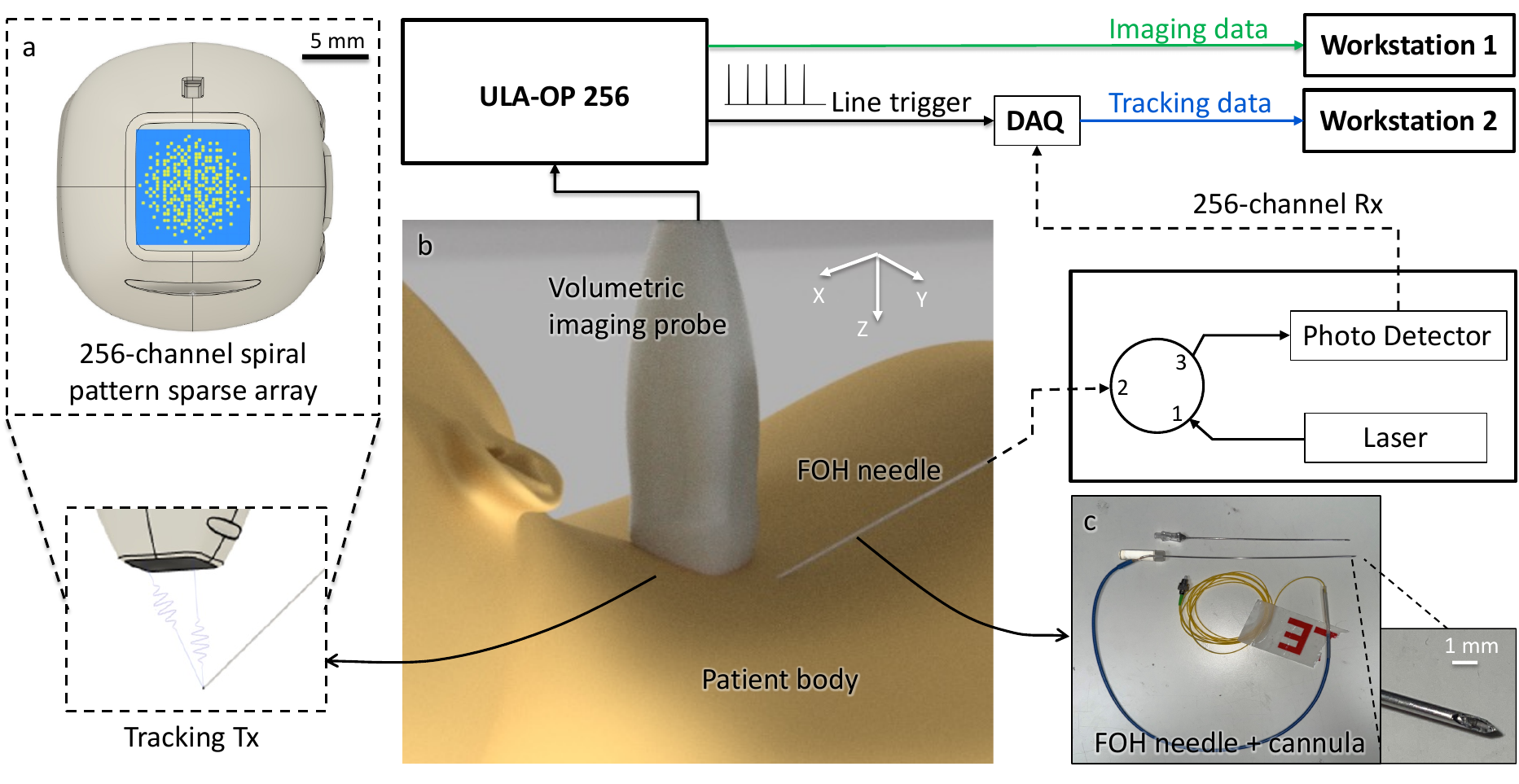}
    \caption{
    Schematic illustration of the 3D needle tracking and volumetric imaging system:  
    (a) 2D matrix array probe with 256 active elements arranged in a spiral-pattern layout, with individual element positions indicated by dots.
    (b) conceptual diagram of the system in a clinical context (e.g., axillary nerve block procedures);  
    (c) 20-gauge needle with an embedded fiber-optic hydrophone (FOH). ULA-OP 256: Ultrasound Advanced Open Platform with 256 channels; DAQ: Data Acquisition;}
    \label{Fig_1}
\end{figure*}

\section{3D Needle Tracking \& Volumetric Imaging System} 
\label{sec_2}
A schematic of the system is shown in Fig.~\ref{Fig_1}. It comprises two main components: (1) a 2D spiral-pattern matrix probe driven by a 256-channel US Advanced Open Platform (ULA-OP 256, MSD Laboratory, Dept. of Information Engineering, University of Florence, Florence, Italy)~\cite{boni2016ula}, and (2) a needle-integrated FOH, from which signals are acquired by a separate data acquisition (DAQ) card. The system enables interleaved volumetric imaging and 3D needle tip tracking. For volumetric imaging, the 2D matrix probe was driven to transmit (Tx) and receive (Rx) plane waves (PWs). For needle tracking, each element of the probe was sequentially excited, transmitting a train of 256 US pulses. For each line trigger, a 4-cycle Gaussian tone-burst was emitted and subsequently detected by the needle-integrated FOH. The complete imaging and tracking data processing pipeline is illustrated in Fig.~S5 (\textbf{Supplementary Material}).

\subsection{Volumetric imaging}
A 2D imaging probe (Vermon S.A., Tours, France) with a \(32 \times 35\) grid of 1024 elements was driven by the ULA-OP 256 and employed to acquire US data. Rows 9, 18, and 27 were disconnected due to routing requirements. The imaging elements have an operational center frequency of 3.7 MHz with a 60\% fractional bandwidth, and the probe features an element pitch of \qtyproduct{300 x 300}{\um}. Only 512 elements were hardwired, following a sparse spiral design based on Fermat's spiral, as described in prior works~\cite{ramalli2015density,peralta20233}. The imaging elements were divided into two groups of 256 elements, each connected to an independent probe connector. On the first connector, the 256 channels were selected using an ungridded, \qty{10.4}{\mm} wide spiral pattern with 256 seeds, with density tapering adjusted via a 50\%-Tukey window (see Fig.~\ref{Fig_1}a). Data were acquired using this configuration. The pulse repetition frequency (PRF) of the ULA-OP 256 was set to 1000 Hz, and radio frequency (RF) data were acquired at a sampling frequency of 26 MHz.

For 3D compounding PW imaging, the probe was programmed to transmit and receive 9 PWs using 4-cycle Gaussian pulses at a center frequency of 3 MHz. The PWs were sequentially beam-steered from $-5^\circ$ to $5^\circ$ in both the lateral (X) and elevational (Y) directions. Two orthogonal slices in the lateral and elevational directions were beamformed using the delay-and-sum (DAS) algorithm and displayed in real-time from the volumetric dataset, as shown in Fig.~S5 (top row). Raw imaging RF data was also saved for offline post-processing Fig.~\ref{Fig_1}).

\begin{figure*}[!ht]
    \centering
    \includegraphics[width=0.95\textwidth]{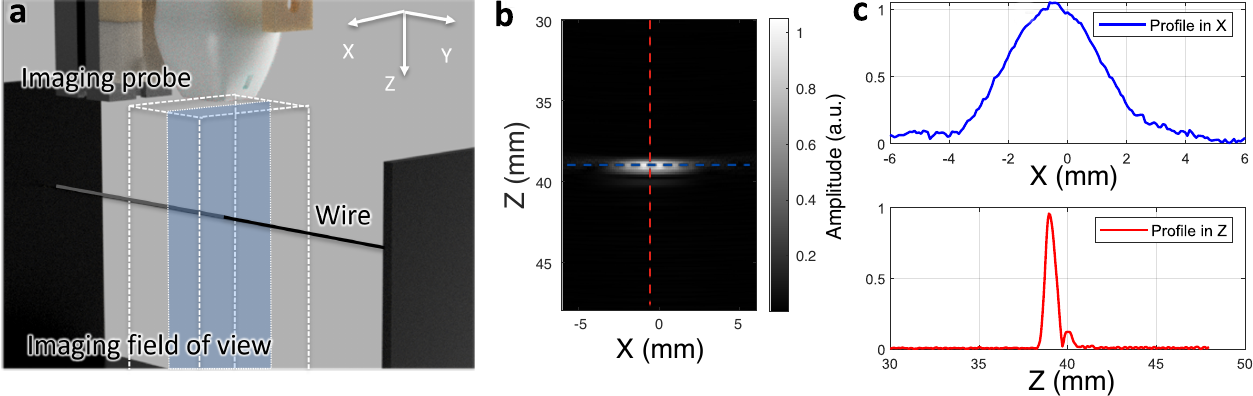}
    \caption{
    Schematic illustration of the volumetric spatial resolution assessment setup and data processing pipeline. (a) Experimental setup; (b) Representative XZ plane slice of a volumetric ultrasound image acquired from a wire phantom; (c) Corresponding point spread function profiles in the lateral (X) and axial (Y) directions.}
    \label{Fig_2}
\end{figure*}

\subsection{3D needle tip tracking}
A FOH, integrated into a 20-gauge, 150~mm-long needle cannula with its tip aligned to the bevel, was used to receive the tracking signal sequences, each consisting of a 4-cycle Gaussian tone-burst pulse. After the needle tip is positioned at the clinical target, both the stylet and FOH are designed to be withdrawn from the cannula. The FOH featured an FP cavity at the distal end of the fiber, consisting of a thin polymer spacer sandwiched between two gold-coated reflective surfaces~\cite{morris2009fabry}. A wavelength-tunable continuous-wave laser (TSL-550, Santec, Japan; 1500--1600\,nm wavelength range) was delivered through the fiber to interrogate the FP cavity. A fiber-optic circulator directed the incident light towards the FP cavity and routed the reflected signal to a photodetector. Acoustic pressure incident on the FP cavity modulates its thickness, thereby altering the reflected optical power, which is measured to extract the acoustic signal~\cite{morris2009fabry}. A DAQ card (M2p.5961-x4, Spectrum Instrumentation GmbH, Germany) was used to acquire the FOH signals via a workstation for offline processing (labeled as 'Workstation 2' in Fig.~\ref{Fig_1}).

The FOH tracking data were processed using the DAS algorithm to reconstruct a 3D tracking image, with the needle tip represented as the sole object as:

\begin{equation}
    S(x, y, z) = \sum_{n=1}^{N} p_n \left( t - \frac{d_n}{c} \right)
    \label{eq_1}
\end{equation}

where:
\begin{itemize}
    \item $S(x, y, z)$ is the computed signal value at the voxel $(x, y, z)$.
    \item $p_n$ is the signal from the $n$-th active transducer element.
    \item $t$ is the time at which the signal is received.
    \item $d_n$ is the distance between the $n$-th element and the voxel at $(x, y, z)$.
    \item $c$ is the speed of sound in the medium.
    \item $N$ is the total number of elements in the transducer array (in this case $N=256$).
\end{itemize}

The center of mass of the reconstructed needle tip region was then calculated to determine the needle tip region. 

\section{Performance evaluation}
\label{sec_3}
\subsection{Volumetric imaging spatial resolution assessment}
To assess the spatial resolution of the imaging system, a point-like imaging target was created by positioning a wire with its axis oriented perpendicular to either the XZ or YZ imaging planes of the probe. This configuration allowed the wire cross-section to appear as a point source in the corresponding 2D imaging slices extracted from the acquired volumetric dataset (Fig.~\ref{Fig_2}b). As shown in Fig.~\ref{Fig_2}a, a nylon wire phantom (\qty{25}{cm} in length and \qty{200}{\micro\meter} in diameter) was used. The phantom was initially submerged in a water tank beneath the imaging probe's field of view, which was mounted vertically and motorized using translational stages (NRT150, Thorlabs, Newton, NJ, USA), with the probe oriented downward into the tank. In the first step, the probe was positioned approximately \qty{10}{mm} above the wire phantom using a translational stage, guided by real-time US imaging. The wire orientation was manually adjusted to align with the probe’s coordinate system. 

For assessing the lateral and axial resolutions, the wire was oriented perpendicular to the XZ imaging plane(Fig.~\ref{Fig_2}b). RF data were acquired with the wire positioned at 4 different depths: 10 mm, 20 mm, 30 mm, and 40 mm. At each depth, the probe was motorized to perform a scan covering a square region spanning X = -3 to 3 mm, and Y = -3 to 3 mm, with 2 mm step increments. This enabled the acquisition of spatial resolution data at various lateral and elevational offsets. At each scan position, data acquisition was repeated five times. The acquired RF data were beamformed, envelope-detected, normalized, and log-compressed. An example XZ plane slice is shown in Fig.~\ref{Fig_2}b. The lateral (X) and axial (Z) resolutions were quantified from the full width at half maximum (FWHM) values of the point spread function (PSF) profiles in the axial and lateral directions extracted at the location of maximum intensity, respectively (Fig.~\ref{Fig_2}c). Similarly, the wire phantom was oriented perpendicular to the YZ imaging plane to assess the spatial resolution in the elevational and axial directions, and the same scanning procedure was repeated.

\subsection{Tracking accuracy assessment}
Tracking accuracy was assessed with the sensorised needle submerged in a water tank and secured in a custom holder attached to the translation stage (Fig.~S6a), which provided the ground-truth needle tip position relative to the US probe. A calibration procedure was conducted using feedback from real-time US imaging to align the scanning stage with the US probe’s coordinate system. As the needle tip was clearly visualized in water with US imaging, this enabled accurate registration of the imaging probe to the 3D scanning stage, ensuring reliable spatial referencing. Full details of this procedure are provided in the \textbf{Supplementary Material}.

For the accuracy assessment, the needle tip followed a predefined trajectory across a 3$\times$3\ horizontal grid, with depths ranging from 10\,mm to 40\,mm. The needle tip was raster-scanned at each depth over a square region spanning X = 0 to 4 mm, and Y = 0 to 4 mm, with 2 mm step increments (Fig.~S6b). FOH data were collected three times at each scan position. The 3D needle tip location was estimated as the center of mass of the needle tip region in the reconstructed tracking image. Tracking accuracy was quantified as the root mean squared errors (RMSEs) of the Euclidean distances between the tracking and known true needle positions, along with the standard deviation (SD) across the three repeated measurements. 

\begin{figure}
    \centering
    \includegraphics[width=0.9\linewidth]{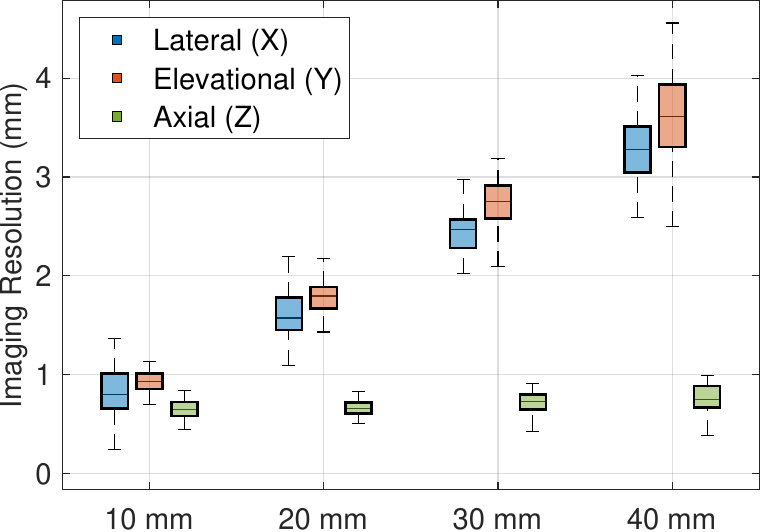}
        \caption{Box-and-whisker plot of volumetric spatial resolution measured using a wire phantom. Measurements were acquired at depths ranging from 10 mm to 40 mm. Each box represents the interquartile range (IQR), with the top and bottom edges corresponding to the 75th and 25th percentiles, respectively; the black line within each box indicates the median. Whiskers extend to the most extreme data points within 1.5 times the IQR from the box edges.}
    \label{Fig_4}
\end{figure}

\begin{figure}[!ht] 
    \centering
    \includegraphics[width=1\linewidth]{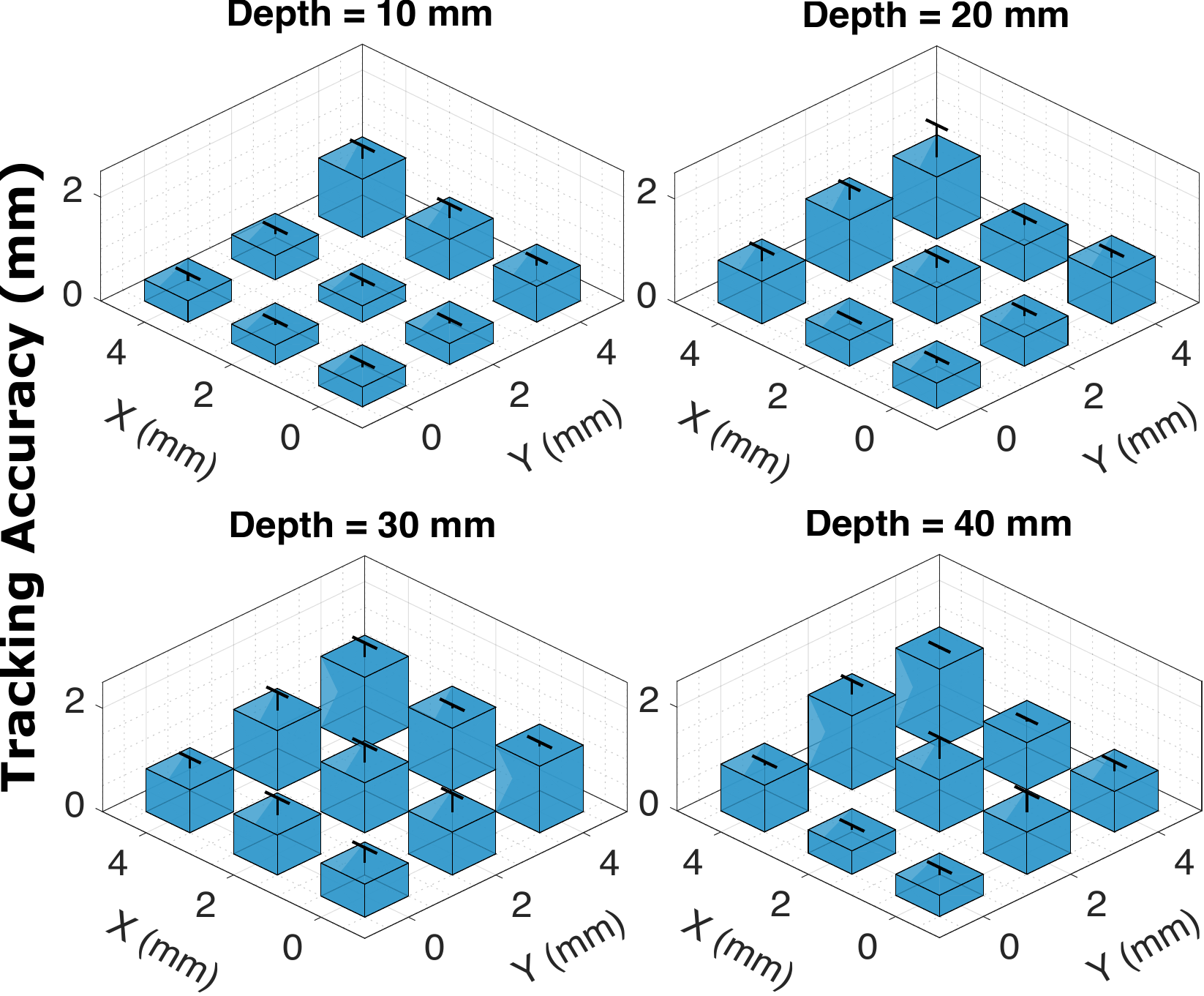}
        \caption{Measured tracking accuracy at varying spatial locations. The error bars indicate standard deviations from 3 repeated measurements.}
    \label{Fig_5}
\end{figure}

\subsection{\textit{Ex vivo} study}
To assess the robustness of tracking under realistic acoustic conditions, needle insertions were conducted at clinically relevant angles using \textit{ex vivo} chicken breast tissue. Needle insertion was guided by an adjustable path that maintained a fixed angle relative to the imaging probe (see Fig.~S8e). The guiding mechanism was constructed using 3D-printed components and incorporated a kinematic rotation joint (Thorlabs, USA). A pre-designed holder, also incorporating a kinematic rotation joint, was used to control the insertion angle. The needle shaft was marked at 2 \,mm intervals to facilitate tracking of insertion depth along the trajectory. Experiments were conducted at three angles relative to the probe surface ($0^\circ$, $30^\circ$, and $60^\circ$), with insertion depths ranging from 10\,mm to 28\,mm. For each insertion, FOH and imaging data were collected at three random positions. Three insertions were performed for each angle. The CNR was calculated from each dataset as:

\begin{equation}
    \text{CNR (dB)} = 20 \log_{10} \left( \frac{|\mu_s - \mu_b|}{\sigma_b} \right)
    \label{eq_2}
\end{equation}
where:
\begin{itemize}
    \item $\mu_s$ is the pixels' mean amplitude of the signal region on the reconstructed image,
    \item $\mu_b$ is the mean amplitude of the background (noise) region,
    \item $\sigma_b$ is the standard deviation of pixel amplitudes in the background region.
\end{itemize}

\subsection{Feasibility study of tracking with a subset of the array elements}
To assess the feasibility of increasing the tracking speed by reducing the number of active elements, different subsets of the spiral array elements from the 2D sparse array probe were created by selecting elements at regular intervals (e.g., every 2nd, 3rd, etc.). This approach maintained the overall spiral configuration while decreasing the total number of transducer elements. DAS beamforming was then used to reconstruct tracking images with these element subsets, with varying transducer counts. The resulting image reconstruction times, CNRs of the reconstructed images, and tracking accuracies were evaluated and compared across the different configurations, based on the data acquired for tracking accuracy assessment in water.

\subsection{Femoral nerve block simulator study}
\begin{figure*} [!ht]
    \centering
    \includegraphics[width=0.8\textwidth]{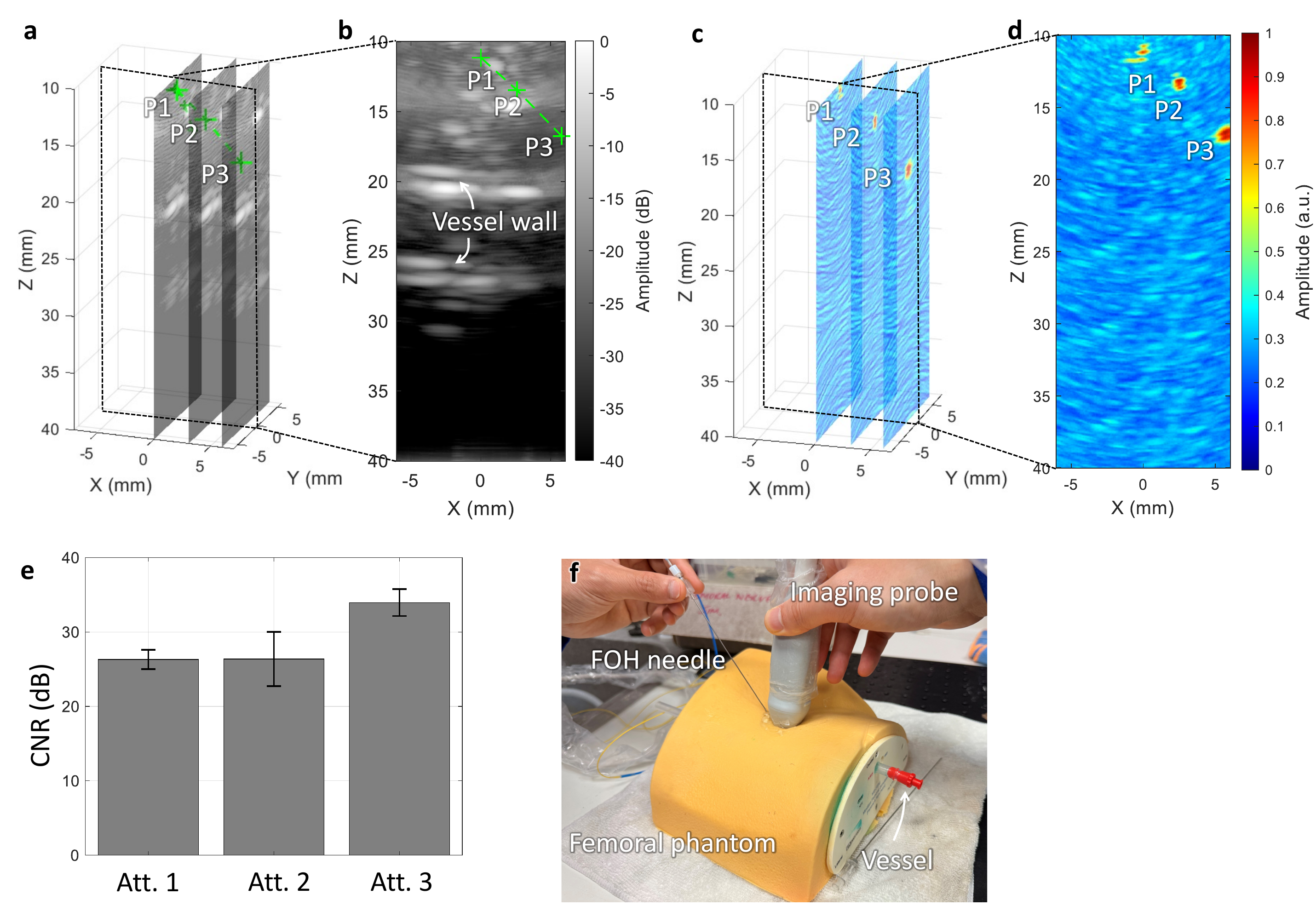}
    \caption{System validation using a femoral nerve block simulator. Representative results from a single needle insertion are shown in (a)-(d). (a) YZ-plane slices from a volumetric ultrasound image; (b) XZ-plane ultrasound image extracted from the volumetric ultrasound data, with its location indicated by the dashed box in (a). Tracked sampling positions (green crosses) are connected by a dashed line to illustrate the needle insertion path, with positions labeled P1, P2, and P3; (c) YZ-plane slices from a volumetric tracking image;
    (d) XZ-plane view of the volumetric tracking image, with its location indicated by the dashed box in (c), showing the tracked sampling positions; (e) Measured contrast-to-noise ratios (CNRs) for all three insertion attempts; Error bars represent standard deviations from 3 sampling locations. (f) Experimental setup. FOH: Fibre-Optic Hydrophone.     
    }
    \label{Fig_6}
\end{figure*}

As a preliminary evaluation of the system’s suitability for clinical translation, needle insertions were performed into a simulator for femoral nerve block procedures (MiniSims, VALKYRIE Simulators, USA). As illustrated in Fig.~\ref{Fig_6}f, the simulator contains a plastic tube approximately \qty{8}{mm} in outer diameter, which is located at an approximate depth of \qty{2}{cm}. To acoustically simulate the femoral artery, the plastic tube was filled with saline water and sealed at both ends to prevent leakage during the experiment.

The FOH needle was inserted towards the femoral artery at three different angles, ranging approximately from \(43.12^\circ\) to \(50.22^\circ\), with insertion depths between \qty{10}{mm} and \qty{25}{mm}. In each case, the needle bevel was consistently oriented upward to enhance acoustic signal reception. For every insertion attempt, both imaging and tracking data were collected at several sampling positions along the needle trajectory. CNR values were calculated for all tracking datasets to evaluate the angular dependence.

\section{Results}
\label{sec_4}
The spatial resolution of the volumetric imaging system is shown in Fig.~\ref{Fig_4}. Over the measurement depth range of \qtyrange{10}{40}{mm}, both lateral and elevational resolutions decreased with increasing depth. Specifically, the lateral resolution varied from $0.87 \pm 0.30$~mm (mean $\pm$ standard deviation) at 10 mm depth to $3.31 \pm 0.37$~mm at 40 mm, while the elevational resolution varied from $0.94 \pm 0.11$~mm to $3.59 \pm 0.37$~mm over the same depth interval. In contrast, the axial resolution remained relatively consistent across all depths, with an average of $0.69 \pm 0.12$~mm. Additional resolution measurements at varying lateral and elevational locations are provided in Fig.~S7 in the \textbf{Supplementary Materials}, illustrating spatial variations in resolution across the field of view.

Fig.~\ref{Fig_5} presents the tracking accuracy measured at various spatial locations within the field of view, offset from the probe center toward the edges, across different imaging depths in water. The positional error (mean $\pm$ standard deviation) increased from $0.40 \pm 0.02$~mm at the center position (0, 0) to $1.30 \pm 0.41$~mm at the peripheral location (4, 4), indicating a gradual reduction in accuracy as the tracking target moved away from the probe center. In addition to lateral variation, tracking accuracy was assessed at multiple depths. The mean tracking accuracy (mean $\pm$ standard deviation) were measured as $0.11 \pm 0.09$~mm, $0.13 \pm 0.11$~mm, $0.17 \pm 0.12$~mm, and $0.20 \pm 0.17$~mm at depths of 10~mm, 20~mm, 30~mm, and 40~mm, respectively. These results demonstrate a consistent degradation in accuracy with increasing imaging depth and distance from the probe center, likely due to the limited viewing angle of the probe and acoustic attenuation.

With the \textit{ex vivo} study, the needle tip was apparent in the reconstructed tracking images at clinically relevant insertion angles ranging from $0^\circ$ to $60^\circ$ over all needle tip positions (Fig.S8, \textbf{Supplementary Materials}). The CNRs of tracking images decreased from $29.99 \pm 8.9\,\mathrm{dB}$ at $0^\circ$ to $9.26 \pm 3.40\,\mathrm{dB}$ at $60^\circ$. This reduction in CNR can be attributed to the diminished sensitivity of the FOH as the needle insertion angle increases relative to the imaging probe.

The feasibility of tracking using a subset of array elements is demonstrated in Fig.~S9 in the \textbf{Supplementary Materials}. As the number of active elements was reduced from 256 to 43, a gradual degradation in image quality was observed. However, this reduction also decreased the FOH data size, resulting in shorter processing times for 3D tracking image reconstruction, accompanied by a decline in image CNR (Fig.~S10).  Despite the reduced number of elements, tracking accuracy remained acceptable, with an error of $0.96 \pm 0.07$~mm when using a 43-element subset (selected at regular intervals of 7), compared to the full 256-element configuration used as reference (Fig.~S10, bottom). In contrast, further reducing the number of elements to 30 (subset regular intervals of 8) resulted in a substantial loss in accuracy (8.11\,mm), which was considered insufficient for reliable tracking. 

With the femoral phantom study, US images were able to provide clear visualization of the femoral artery; however, the needle was not visible (Fig.~\ref{Fig_6}a and \ref{Fig_6}b). In contrast, the needle tip position was successfully tracked and visualized using the reconstructed tracking images, as illustrated in Fig.~\ref{Fig_6}c and \ref{Fig_6}d. The CNRs of the tracking images across all sampling positions ranged from \qtyrange{26.3}{33.9}{dB}, indicating robust localization performance (Fig.~\ref{Fig_6}e). These results highlight the complementary roles of the two modalities: US imaging effectively identifies anatomical structures such as vessels, while the tracking capability provides accurate and consistent localization of the needle tip, particularly in cases where the needle is not visible in the US image.

\section{Discussion}
\label{sec_5}
Accurate localization of interventional devices such as medical needles relative to target anatomy remains a long-standing challenge for safe and effective guidance in US-guided procedures. Previous studies have typically focused on either 3D needle tracking or 3D US imaging alone. To the best of our knowledge, this work presents the first integrated system capable of real-time volumetric imaging and 3D needle tracking using a 2D sparse matrix array imaging probe combined with fiber-optic US sensing. This could enable clinicians to visualize surrounding 3D anatomy simultaneously and accurately track the needle tip in real time. Based on evaluations in water, \textit{ex vivo} tissue, and femoral nerve block phantoms, the system demonstrated reliable tracking and imaging performance, including effective needle tip 3D localization at steep insertion angles, highlighting its potential for future clinical translation.

% discussion on volumetric imaging: 
For volumetric imaging, the use of a 2D spiral-pattern matrix probe enables higher imaging speeds than mechanically scanned probes, while also reducing channel count, transducer cost, and computational burden~\cite{peralta20233}. In water-based validation experiments, the system achieved overall spatial imaging resolutions (mean~$\pm$~standard deviation) of $1.67 \pm 0.22$\,mm, across a depth range of \qtyrange{10}{40}{mm}. Procedure targets, such as peripheral nerves, typically range in diameter from 2 to 8\,mm~\cite{guay2019use}, and as such, achieving sub-millimeter resolution is essential to accurately identify nerve structures and guide needle placement using volumetric US imaging. Spatial resolution can be improved by employing higher-frequency probes, as clinical US imaging probes used in nerve block procedures typically operate at frequencies up to 15\,MHz, with some models reaching 20\,MHz. However, increasing the US frequency also results in larger data volumes and higher system costs. Another promising solution to enhance imaging quality is to increase the number of PW transmissions, thereby enabling finer beam steering and improved volumetric image reconstruction. However, this comes at the cost of increased data load and reduced frame rate due to the additional number of transmissions. Alternatively, coherent multi-transducer ultrasound imaging offers a compelling approach. By combining data from multiple probes to synthetically extend the effective aperture, this technique enhances both spatial resolution and field-of-view without compromising the frame rate~\cite{peralta2019coherent}.

% discussion on 3D tracking: 
For 3D needle tracking, the FOH offers several advantages over conventional piezoelectric transducers. It is more cost-effective than miniaturized piezoelectric sensors, and its ultra-compact diameter of \qty{150}{\micro\meter} enables seamless integration into a wide range of medical needles without obstructing the lumen, thus preserving the ability to deliver medication. The FOH also exhibits a broad frequency bandwidth (~1--20\,MHz), allowing compatibility with clinical ultrasound transducers with varying frequencies, unlike many piezoelectric elements, which typically exhibit narrower bandwidths~\cite{morris2009fabry}. In addition, its miniature size contributes to improved omnidirectional sensitivity, enhancing spatial tracking robustness~\cite{zhang2011miniature}. Importantly, the FOH is inherently immune to electromagnetic interference, making it suitable for integration into MRI-guided procedures.

%tracking and imaging frame-rate
Currently, tracking data are acquired in real time but processed offline on a local workstation (GPU: Quadro RTX 5000), before being overlaid onto the corresponding ultrasound image frames. Each image-tracking cycle involves 265 trigger events generated by the ULA-OP 256, operating at a PRF of 1000~Hz, resulting in an effective frame rate of approximately 3 frames per second. To further increase the tracking frame rate, one potential approach is to decrease the number of active elements on the probe following the spiral pattern layout. A preliminary investigation was carried out by sparsifying the active element configuration within this pattern. Experimental results in water demonstrated that accurate 3D needle tracking could be achieved using as few as 43 elements. This sparsified configuration has the potential to reduce the tracking frame rate by up to a factor of five.

The tracking image reconstruction was performed using a volumetric DAS algorithm implemented in MATLAB for research purposes. Under the current computational setup, processing each frame of 256-channel tracking data requires an average of 22.79~seconds, which limits the achievable frame rate. Reducing the computational time is essential to enabling real-time intraoperative use. Potential strategies to address this limitation include adopting parallel computing approaches such as GPU acceleration and reducing the reconstruction volume to focus only on the region containing the needle tip. These improvements could significantly accelerate processing and enhance the feasibility of clinical implementation.

\section{Conclusion}
\label{sec_6}
We have developed a 3D ultrasonic needle tracking and real-time volumetric imaging system by integrating a 2D spiral-pattern matrix array with an FOH-equipped needle. The system demonstrated spatial resolution ranging from 0.69 to 2.26 mm and sub-millimeter tracking accuracy. Its clinical feasibility was validated using a nerve block training phantom, illustrating its capability for simultaneous anatomical visualization and accurate needle tip localization. The proposed approach shows strong potential to improve the precision and safety of interventional procedures by providing real-time 3D imaging and guidance.

\appendix
\textbf{Supplementary Materials} associated with this article can be found in the online version.

\bibliographystyle{IEEEtran}
\bibliography{citation}



\section*{Supplementary material}

:

## Supplementary Material for

*Volumetric ultrasound imaging with a sparse matrix array and integrated fiber-optic sensing for robust needle tracking in interventional procedures*

**Authors**: Weidong Liang[1,*], Javad Rostami[1,*], Christian Baker[1], Simeon West[2], Athanasios Diamantopoulos[3], Sunish Mathews[4], Adrien E. Desjardins[5], Sebastien Ourselin[1], Laura Peralta[1], and Wenfeng Xia[1,†]

Affiliations:

[*] These authors contributed equally to this work.

[1] School of Biomedical Engineering & Imaging Sciences, King's College London, London, United Kingdom.

[2] Department of Anaesthesia, University College London Hospital, London, United Kingdom.

[3] Department of Interventional Radiology, Guy's & St Thomas' NHS Foundation Trust, London, United Kingdom.

[4] School of Science and Engineering, University of Dundee, Dundee, United Kingdom.

[5] Department of Electrical and Computer Engineering. The University of British Columbia (UBC), Vancouver, Canada.

[†] Corresponding authors: Wenfeng Xia: wenfeng.xia@kcl.ac.uk

**One-sentence Summary**: This document presents the procedures used to align the needle motion with the imaging probe's coordinate system and corresponding Supplementary Figures to the journal article.

**Calibration procedure for aligning the scanning stage with the ultrasound probe's coordinate system:**

Tracking accuracy was evaluated using a translation stage that provided the ground-truth position of the needle tip relative to the ultrasound probe. To align the scanning stage with the coordinate system of the ultrasound probe, a calibration procedure was performed using feedback from real-time ultrasound imaging. Since the needle tip was clearly visible in water during ultrasound imaging, this allowed for precise registration of the imaging probe to the 3D scanning stage, ensuring consistent and reliable spatial referencing.

**Experimental setup:** The imaging probe was mounted on a probe clamp with adjustable friction-based joints. As shown in **Fig. S1**, the needle was secured on a mechanical holder that was mounted on a translational stage, which allowed motion in three orthogonal directions.

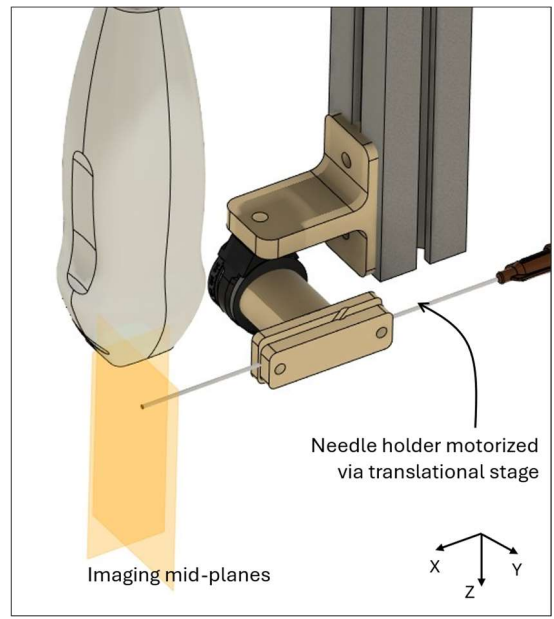

**Fig. S1**: Illustration of the experimental setup for tracking accuracy assessment in a water tank. The probe is mounted facing downward using a mechanical clamp, while the needle is held by a custom holder and motorized using a translational stage.

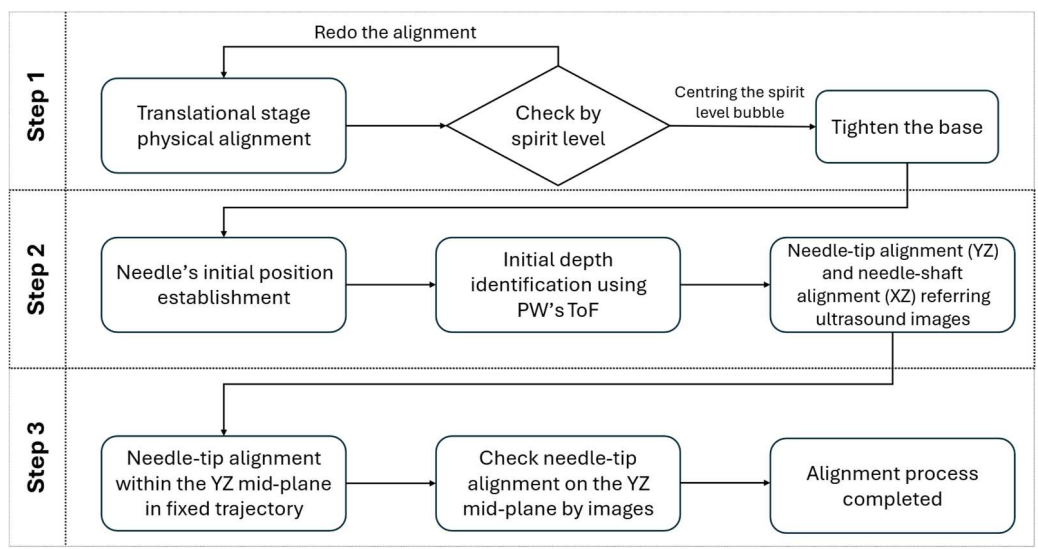

**Fig. S2**: The flowchart of the needle alignment procedure. PW: Plane-wave; ToF: Time-of-flight.

**Calibration procedure:**

**Step 1: physical alignment of the translation stage**

The alignment process begins with the mechanical setup. The imaging probe, mounted on a fixed holder, is manually aligned with the translational stage axes. To ensure proper orientation, the translational stage is first aligned with the experimental bench. A spirit level is used to verify that the axial (Z) stage rail is vertically upright, while the lateral (X) and elevational (Y) stages are horizontally levelled. Alignment is confirmed when the spirit level bubble is centered, after which the base is securely fixed in place.

**Step 2: establishing the needle's initial spatial position relative to the imaging probe**

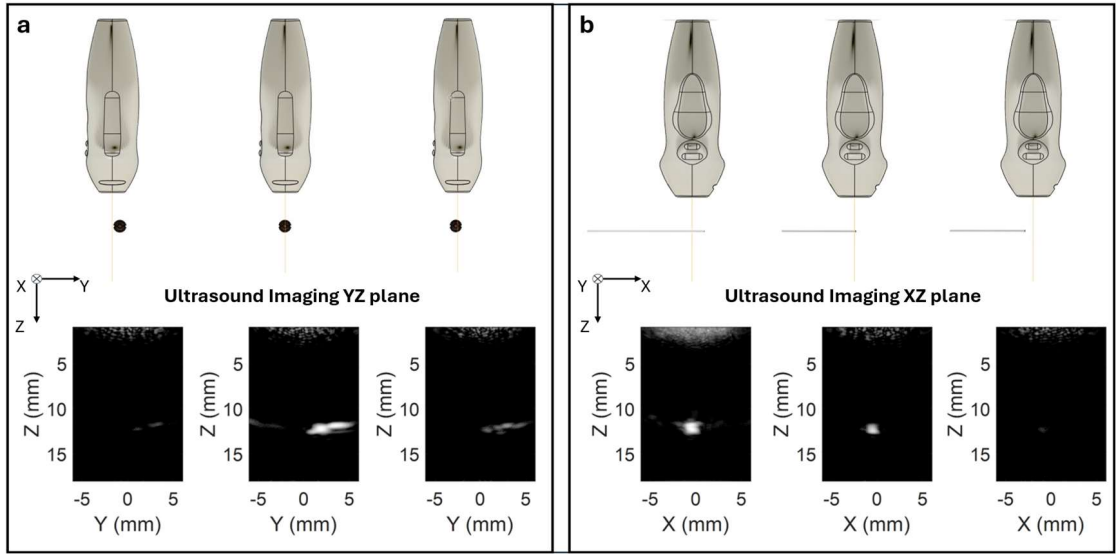

**Fig. S3:** (a) Needle shaft alignment in the YZ mid-plane. The top row shows three needle positions; the bottom row shows corresponding ultrasound (US) images. Left: The needle is barely visible, indicating it is outside the imaging plane. Middle: A bright line appears, confirming alignment with the YZ mid-plane. Right: The brightness fades as the needle moves off-plane. (b) Needle tip alignment in the XZ mid-plane. The top row shows 3D views; the bottom row shows corresponding US images. Left: The needle tip crosses the XZ plane, visible as a bright spot. Middle: Gradual retraction along the Y-axis dims the spot, indicating tip alignment. Right: The spot disappears, showing the tip is off-plane.

To transform needle motion from the translational stage coordinate system to the imaging coordinate system, it is essential to determine a set of known, corresponding positions in both frames. The alignment process begins by positioning the needle tip at the center of the imaging field-of-view (FoV), defined by the intersection of the lateral (X) and elevational (Y) centering planes. This reference position was approximately 10 mm from the probe surface. At this location, the needle, mounted on the needle holder, was oriented along the lateral (X) axis of the imaging coordinate system, as shown in **Fig. S1**.

First, the depth of the needle tip was adjusted to approximately 10 mm. The 2D matrix probe was programmed to transmit plane waves (PWs) along the axial direction, and the true axial position of the needle tip was estimated based on the **time-of-flight** of the signals received by the FOH in water at room-temperature.

The lateral and elevational position of the needle tip was adjusted to lie at the center of the XY imaging plane. The imaging probe provides real-time imaging in two orthogonal cross-sectional planes, XZ and YZ, which intersect at the mid-plane of the imaging coordinate system, as

shown in **Fig. S1**. Imaging feedback was used to guide the adjustment of the needle's initial position. Lateral alignment along the X-direction was then performed by adjusting the needle position until the shaft appeared as a bright line in the YZ mid-plane view (**Fig. S3a**). Elevational alignment was achieved by translating the needle until its tip appeared as a bright spot in the real-time XZ mid-plane view (**Fig. S3b**). At this stage, the initial position of the needle tip was established in both the imaging and translational stage coordinate systems.

**Step 3: validation of needle trajectory alignment using iterative procedures**

Once the needle's initial position is preliminarily determined, additional calibration is required to align the needle's motion with the imaging coordinate axes. As shown in Fig. **S4i** the needle tip was calibrated to move within the YZ imaging mid-plane. In the first step, the needle was moved axially from its initial position (denoted as position 'A') to position 'B', with a displacement of 10 mm. The imaging probe was then manually adjusted to ensure the needle tip just appeared in the YZ imaging mid-plane. To verify proper alignment, the needle was also motorized laterally, as shown in **Fig. S4ii**, to confirm consistent tip visibility. The needle was then sequentially moved through the remaining predefined positions (A → B → C → D → E → F), and the same calibration steps were repeated at each point. At each position (A to F), it was necessary to confirm that the needle tip was accurately aligned within the YZ imaging mid-plane. This was first verified by visualizing the needle tip in the real-time YZ image. The tip was then slightly translated along the X-direction; if it disappeared immediately from view, this indicated correct alignment within the mid-plane. If the tip remained visible, the probe orientation was manually adjusted until the tip just disappeared from the image. This alignment procedure was repeated iteratively for each position, with each iteration followed by the same verification steps to ensure accurate registration and consistent needle tip visibility. Once the optimal probe orientation was achieved, the true needle position relative to the probe's coordinate system was established.

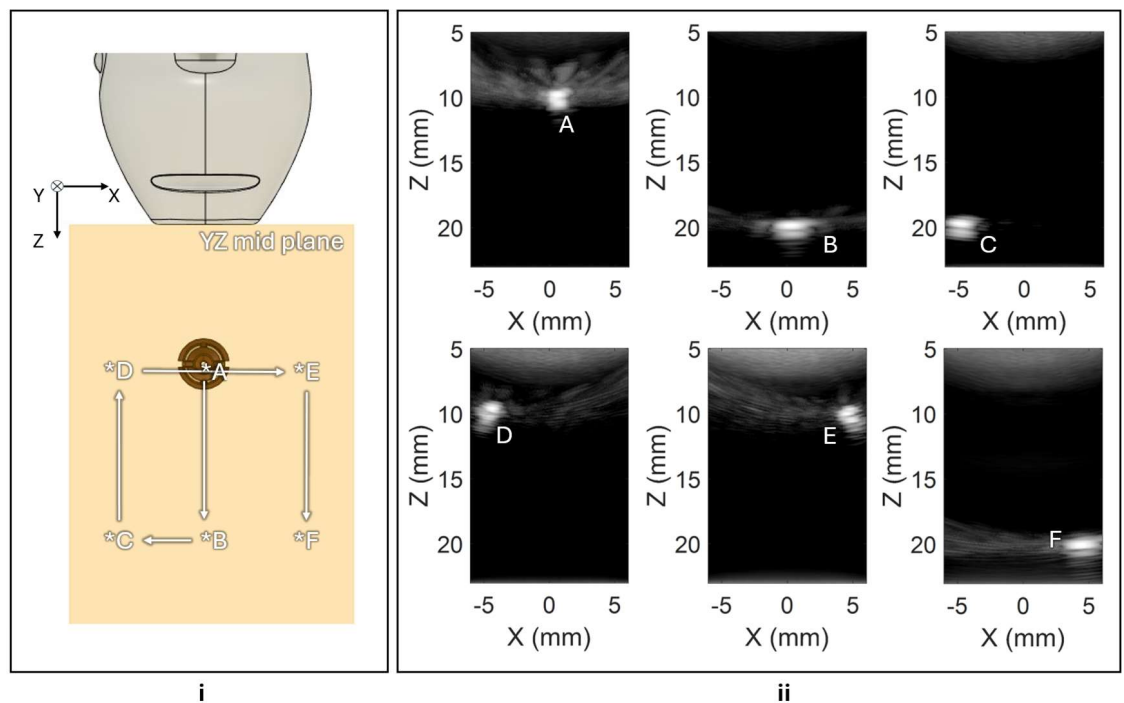

**Fig. S4:** (i) shows the needle trajectory in YZ mid-plane. (ii) shows the needle-tip in the YZ imaging mid-plane in all testing positions.

**Summary**

This document outlines the procedures used to align the needle's motion—controlled by the translational stage—with the coordinate system of the ultrasound imaging probe. Accurate alignment is essential for determining the true position of the needle tip during water-tank accuracy assessment. The process involved initial mechanical setup, positioning the needle tip at the center of the imaging field-of-view, and iterative adjustments based on real-time imaging feedback. By establishing corresponding reference points in both coordinate systems, the needle's motion could be accurately mapped into the imaging domain. This alignment is a key step toward reliable accuracy evaluation of the tracking system.

**Supplementary Figures**

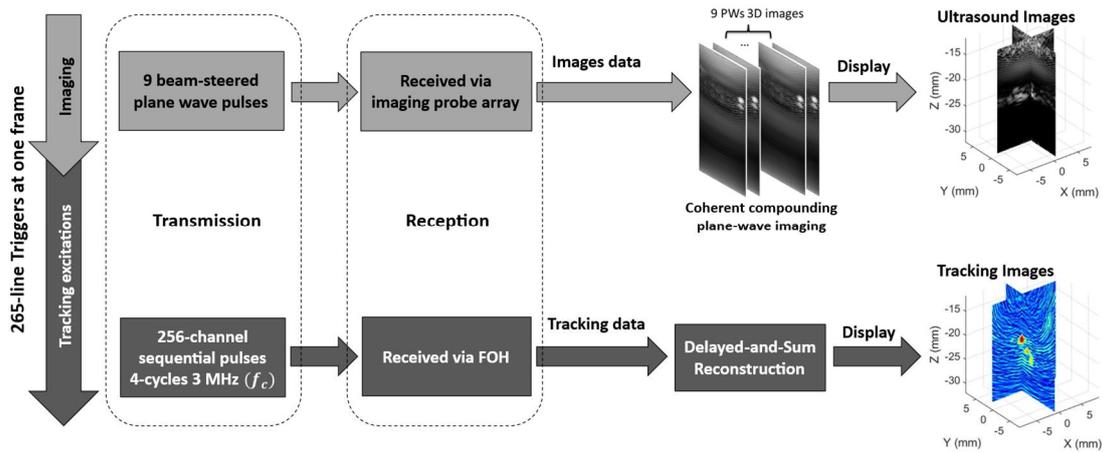

**Fig. S5**: Flowchart of the imaging and tracking signal processing pipeline. Top: volumetric ultrasound data acquisition and real-time imaging in two perpendicular planes; bottom: 3D needle tracking; FOH: Fiber-Optic Hydrophone; PWs: Plane-waves.

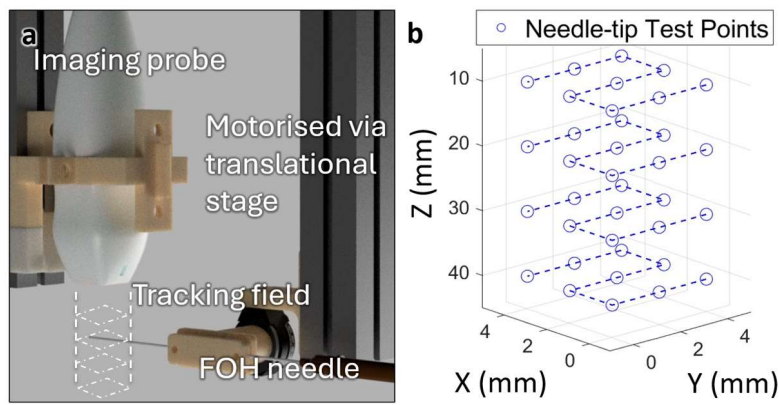

**Fig. S6:** 3D needle tracking accuracy assessment (in water): (a) Experimental setup; (b) Needle motion trajectory in water. FOH: Fiber-Optic Hydrophone.

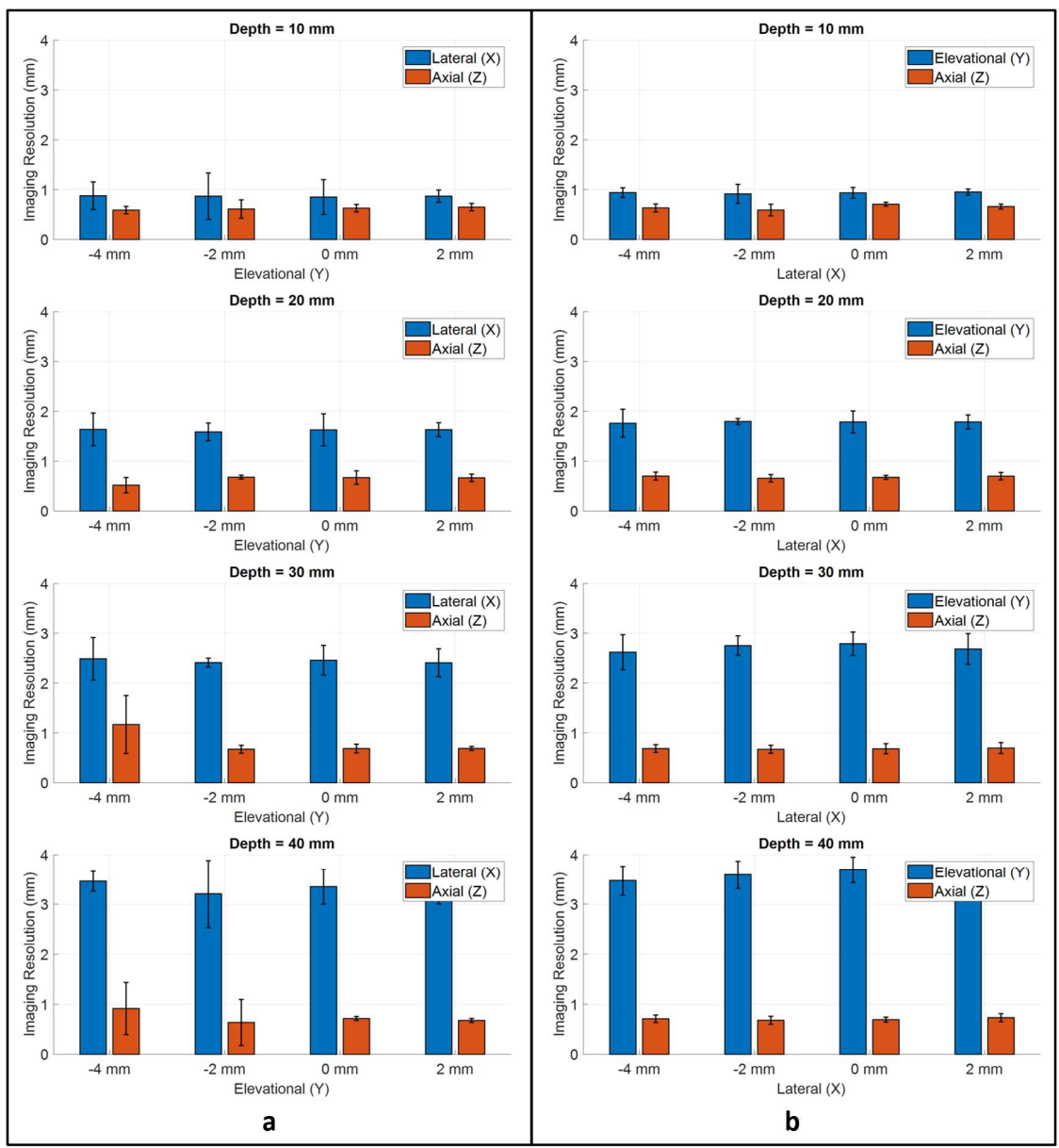

**Fig. S7**: Spatial imaging resolution results at varying lateral and elevational offsets across depths of 10 – 40 mm. (a) Resolution assessed from XZ-plane images with offsets along the Y-direction; (b) resolution assessed from YZ-plane images with offsets along the X-direction.

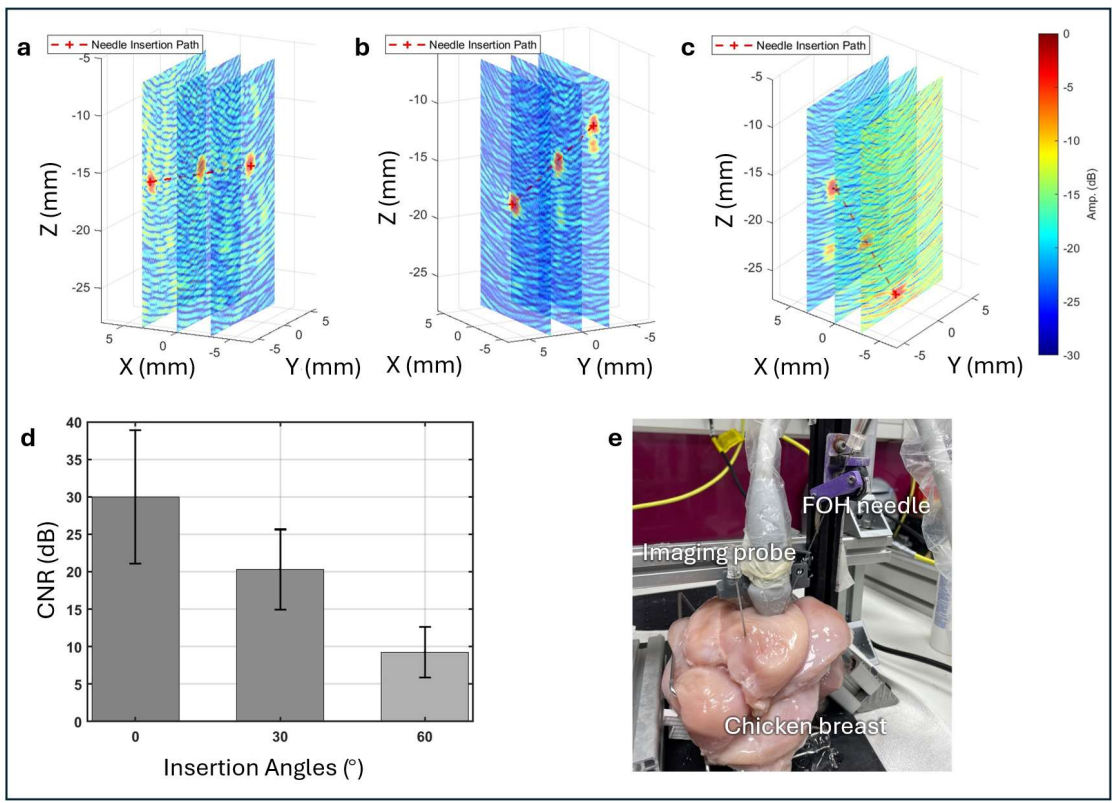

**Fig. S8**: Tracking Contrast-to-noise ratio （CNR） assessment in an ex vivo tissue study. (a–c) XZ-plane reconstructed needle-tip image slices from volumetric data corresponding to insertion angles of 0°, 30° and 60°. (d) CNR of the tracked needle-tip positions across insertion angles.; (e) experimental setup. FOH: Fiber-Optic Hydrophone.

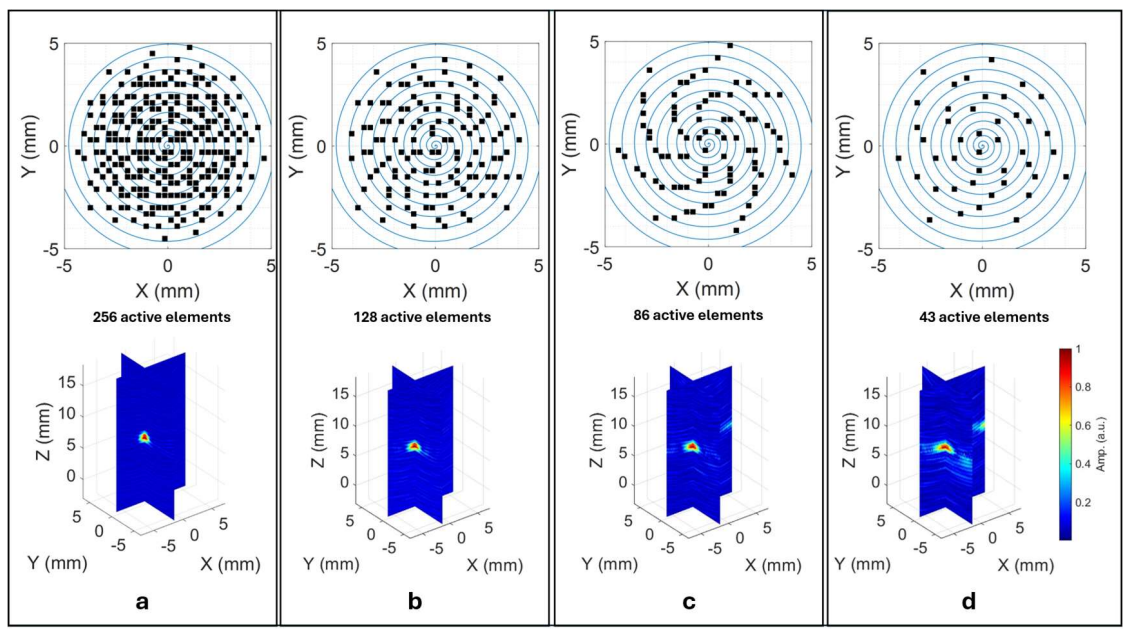

**Fig. S9**: Tracking performance using a downsized spiral-pattern array (dotted-line layout), with the number of active elements reduced from 256 to 43. (a--d) Top row: selected active elements within the spiral array for each configuration; bottom row: corresponding reconstructed orthogonal tracking planes.

Submitted to IEEE Transactions on Biomedical Engineering

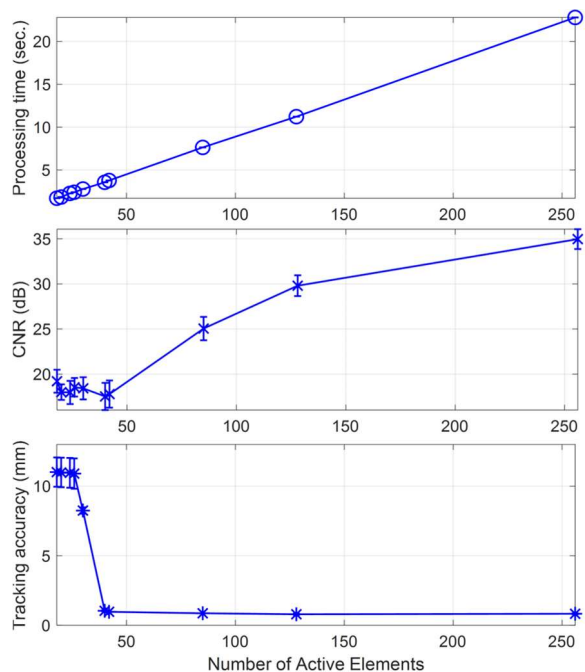

**Fig. S10:** Tracking performance in water using varying numbers of active elements. Top: processing time per frame; middle: contrast-to-noise ratio (CNR) of the tracking images; bottom: tracking accuracy quantified by root-mean-square error (RMSE).



% \includepdf[
%   pages=-,
%   pagecommand={},
%   fitpaper=true,
%   addtotoc={1,section,1,Supplementary Material,suppl}
% ]{TBME_suppl.pdf}

\end{document}